\documentstyle[aps,epsfig,epsf]{revtex}

\begin{document}

\twocolumn[
\hsize\textwidth\columnwidth\hsize\csname @twocolumnfalse\endcsname

\title{Interface and contact line motion in a two phase fluid under shear flow}
\author{Hsuan-Yi Chen and David Jasnow}
\address{Department of Physics and Astronomy, University of Pittsburgh,
Pittsburgh, Pennsylvania 15260}
\author{Jorge Vi\~nals}
\address{Supercomputer Computations Research Institute, Florida State 
University, Tallahassee, Florida 32306-4130,
and Department of Chemical Engineering, FAMU-FSU College of Engineering,
Tallahassee, Florida 32310-6046}
\date{\today}
\maketitle

\begin{abstract}
We use a coarse grained description to study the steady state interfacial
configuration of a two phase fluid under steady shear. Dissipative
relaxation of the order parameter leads to interfacial slip at the contact
line, even with no-slip boundary conditions on the fluid velocity. 
This relaxation occurs within a characteristic length scale $l_{0}=
\sqrt{\xi D/V_{0}}$, with $\xi $ the (microscopic) interfacial thickness, $D$
an order parameter diffusivity, and $V_{0}$ the boundary velocity. The
steady state interfacial configuration is shown to satisfy a scaling form
involving the ratio $l_{0}/L$, where $L$ is the width of the fluid layer,
for a passive interface, and the capillary number as well for an active
interface.
\end{abstract}

\pacs{68.10.-m,83.10.Lk}

\narrowtext
]

We reexamine a classical hydrodynamic problem in which a two-phase fluid
is placed in a uniform shear flow by displacing two infinite, parallel
planar boundaries at constant relative velocity. Instead of considering
the case of two immiscible fluids separated by an infinitely sharp 
interface, our analysis involves two coexisting fluid phases.
The interface, which is
diffuse even without shear, is stretched by the flow, but not
indefinitely as interfacial slip relieves the increasing shear stress at the
three phase junction. A mesoscopic description of the two-phase fluid is
introduced to show that diffusive relaxation of the order parameter (e.g.,
concentration) necessarily contributes to macroscopic 
{\em interfacial slip}. Since in typical situations mutual diffusion occurs 
on lengths large
compared to microscopic (molecular) lengths, it is not necessary to
introduce microscopic {\em slip of the fluid velocity} near the contact
lines. Finally, we also show that in the steady state, the interfacial 
configuration satisfies a scaling form, provided that 
the interfacial curvature is small compared to
its inverse thickness.

The classical analysis of the contact line profile and motion in partially
wetting fluids at low capillary numbers is due to Cox \cite{re:cox86}, and
Dussan and Davis \cite{re:dussan86}. A matched asymptotic expansion
describes both the interfacial configuration and flow field by considering
three separate regions. Close to the boundary, the fluid is allowed to slip
by invoking a phenomenological relation between the slip velocity and the
local shear stress. This relation can also be viewed as introducing a slip
length of microscopic dimensions and an effective slip boundary condition
for the velocity. Far from the boundaries, capillary forces are
balanced by viscous stresses. An intermediate region allows matching between
the two. Reviews of this treatment have been given in \cite
{re:degennes85,re:oron97}. Early experiments by Dussan \cite{re:dussan89}
confirmed this picture. More recent experiments by Lichter 
\cite{re:lichter99} have addressed the range of capillary numbers within
which this solution is a good approximation near the contact line.

If the fluid in question is a binary at coexistence, interfacial deformation
induces concentration gradients through a Gibbs-Thomson 
relation~\cite{Bray}.
The resulting diffusive transport leads to effective interfacial
slip at the contact line even when the velocity strictly satisfies a
no-slip boundary condition. Increasing interfacial curvature leads to
increasing diffusive transport which eventually balances advection by the
flow. A steady state can be maintained. Within a coarse-grained or
mesoscopic description of the two phase fluid \cite{HH;OnuK}, this balance
occurs within a characteristic length scale $l_{0}$, which is 
the geometric mean of the interfacial thickness and the diffusion length, 
and hence is a small quantity although not of molecular dimensions.
Dissipation within
this scale effectively prevents the formation of stress singularities at the
contact line. A mesoscopic description of contact line motion has already
been given \cite{re:seppecher96;re:jacqmin99}, by using the same model
equations which we describe below. These authors, however, focused on the
limit $l_{0}\rightarrow 0$ (in our notation), which corresponds to effective
immiscibility on all scales. The effect that we discuss, however, requires
consideration of distances to the boundary small compared to $l_0.$ 

We consider an incompressible binary mixture in two spatial dimensions
confined between two planar boundaries of infinite extent located at $x=\pm
L/2$. The bounding planes are displaced with constant
velocity $\pm V_{0}/2\ \hat{{\bf y}}$. 
We study the steady state interface configuration in terms of $V_{0}$, $L$,
and the physical parameters of the fluid. Within a coarse grained
description of a binary fluid~\cite{J&V,BJO,re:gurtin96,re:anderson98},
the order parameter $\phi$, e.g., concentration, satisfies that 
$ \phi =\phi _{+}>0$ for one phase, and $\phi =\phi _{-}<0$ for the other. The
nominal position of the interface is defined as the locus 
$ \phi =0$. The temporal evolution of the order parameter is governed by 
\begin{eqnarray}
\frac{\partial \phi }{\partial t}+{\bf v}\cdot {\bf \nabla }\phi =M
\nabla^{2}\mu \ ,  \label{Cahn-Hill}
\end{eqnarray}
where ${\bf v}$ is the fluid velocity, the chemical potential $\mu
\lbrack \phi ]=\delta F/\delta \phi ,$ with $F[\phi ]$ the coarse
grained free energy functional, and $\delta /\delta \phi $ stands for
variational differentiation. $M$ is a mobility coefficient assumed to be
constant. The velocity ${\bf v}$ obeys a modified Navier-Stokes equation, 
\begin{eqnarray}
\rho \left( \frac{\partial {\bf v}}{\partial t}+{\bf v}\cdot {\bf \nabla v}
\right) =-{\bf \nabla }p+\mu {\bf \nabla }\phi +\eta \nabla ^{2}{\bf v}\ ,
\label{Navier-Stokes}
\end{eqnarray}
where $\rho $ and $\eta $ are the density and shear viscosity of the fluid,
assumed to be independent of $\phi $, and $p$ is the pressure. The second
term on the right hand side of Eq.~(\ref{Navier-Stokes}) incorporates the
effects of capillarity. We choose the standard free energy 
$F[\phi ]=\int dV\left( \frac{K}{2}|\nabla \phi |^{2}+f(\phi
)\right)$, with $K$ constant. The first term represents the excess free
energy due to spatial inhomogeneities of $\phi$, and $f(\phi )$ is the local
part of the free energy density. Generalizations involving, say, higher
gradients or forcing due to short-ranged wall interactions can be
incorporated.

Several macroscopic physical quantities follow from this coarse grained
description ~\cite{DJ,Bray}. The surface tension is given by $\sigma
=\int_{-\infty }^{\infty }K \left[ \partial \phi _0 /\partial y\right]^{2}
dy$, where $\phi_0$ is the equilibrium order parameter profile
at coexistence (we take $\mu(\phi_{0}) = 0$ in what follows). Furthermore,
the chemical potential at a gently curved interface with radius of curvature 
$R_{c}$, relative to its value at coexistence, is $\mu =-\sigma /R_{c}\Delta
\phi$, where $\Delta \phi =\phi _{+}-\phi _{-}$ is the miscibility gap. 
Another important quantity is the (finite) interfacial thickness which is 
proportional to $\xi =\sqrt{K \chi}$ where $\chi^{-1} = \left( 
\partial^{2}f/\partial \phi^{2}\right)_{\phi_{\pm}}$. Since we assume that 
the fluid is not too close to criticality, $\xi$ is of molecular 
dimensions. Finally, we focus exclusively on moderate shear rates, such that 
$\xi /R_{c}\ll 1$, a limit commonly realized in practice. 

We begin by illustrating the effect of the shear on the interface
through dimensional analysis. If the interfacial mean curvature is
$1/l$, then the local deviation of $\mu$ from coexistence is proportional to 
$1/l$. We therefore choose the following
dimensionless (barred) quantities: $\mu =\frac{\sigma }{l\Delta \phi }\ 
\bar{ \mu}$, $\phi =(\Delta \phi )\bar{\phi}$, and ${\bf v}=V_{0}\ 
{\bf \bar{v}}$. In steady state, 
$Pe{\bf \bar{v}}\cdot \bar{\nabla} \bar{\phi}=\bar{\nabla}^{2}
\bar{\mu}$, where $Pe=V_{0}\Delta \phi^{2}l^{2}/M\sigma $. On
small scales ($l\sim \xi),\ Pe$ becomes very small and hence $\bar{\nabla}^{2}
\bar{\mu}=0$ defines the equilibrium order
parameter profile in the absence of shear. Diffusion and miscibility are
important. If, on the other hand, $l\sim L\rightarrow \infty $, then 
${\bar{\nabla}}\bar{\phi} \cdot {\bf \bar{v}}=0$, corresponding either to a
homogeneous configuration, or to an interface parallel to the flow.
Miscibility effects are not important at this scale. There
exists, however, an intermediate length scale $l_{0}=\sqrt{ M\sigma
/V_{0}\Delta \phi ^{2}}$ for which $Pe=1$. Advection now balances order
parameter diffusion along the interface driven by chemical potential 
gradients; this is the length scale of interest. Given the definition of the
interfacial thickness $\xi $, we can write $l_{0}\propto \sqrt{\xi D/V_{0}}$
, where $D=M/\chi$ is a diffusion coefficient. Hence this new length scale
is the geometric mean of the interfacial thickness and a diffusion length 
$ D/V_{0}$. For a typical system, $D\sim 10^{-5}cm^{2}/s$, $\xi \sim 10\ \AA,$ 
and for a wall velocity $V_0 \sim 1~cm/s$, $l_{0}\approx 10 \ nm$. 

We next present an approximate analytic treatment that is valid for length
scales larger than $\xi$. We find a scaling solution for the steady state
interfacial profile for both passive interfaces (neglecting capillarity
induced backflow, i.e., model B in the critical dynamics lexicon) and active
interfaces. We then present a numerical solution of the full set of
governing equations to validate the scaling forms derived, at least within
the range of parameters which is numerically accessible.

For a passive interface, the velocity field is given by 
$ {\bf v_s}({\bf r}) = x V_0/L \ \hat{{\bf y}}$. Equation
(\ref{Cahn-Hill}) in steady state is written in integral form, 
\begin{eqnarray}
M \mu({\bf r}) = \int dr^{\prime}G({\bf r, r^{\prime}}) \left[{\bf v} \cdot 
{\bf \nabla}\phi \right]_{r^{\prime}} \ ,  \label{inverse-Cahn-Hill}
\end{eqnarray}
where $G({\bf r, r^{\prime}})$ is the Green's function of the Laplacian
operator satisfying Neumann boundary conditions at $x={\pm}L/2$ \cite{green}
. In two dimensions the Green's function is dimensionless. Let the position
of the nominal interface be ${\bf r}_I = y_I(x) \hat{{\bf y}} + x \hat{{\bf x
}}$, where $\hat{{\bf x}}$ and $\hat{{\bf y}}$ are unit vectors. When radii
of curvature are large compared to $\xi$, one can make the standard
approximation, ${\bf \nabla} \phi \simeq (\Delta \phi) \delta({\bf r}-{\bf 
r}_{I}) \hat{{\bf n}}$, and $\phi = \phi_\pm$.~\cite{OK} Using now standard
methods (see, e.g., Ref.~\cite{Bray,OK}), one multiplies Eq.~(\ref
{inverse-Cahn-Hill}) by $\partial \phi/\partial g$ ($g={\bf r}\cdot 
\vec{\nabla} \phi/|\nabla \phi|$) and integrates $g$ across the interface to
derive an equation that depends on the coordinates of the interface alone.
The result is 
\begin{eqnarray}
\frac{1}{R(X)} = - \left(\frac{l_0}{L}\right)^{-2} \int^{1/2} _{-1/2}
dX^{\prime}X^{\prime}G({\bf R_I},{\bf R_I^{\prime}}) \ ,  \label{modelB}
\end{eqnarray}
where we have now scaled lengths by $L$ ($R = R_c/L$, $X = x/L$, $Y = y/L$),
and ${\bf R_I}(X)={\bf r}_I(x)/L$ is the scaled location of the interface.
This equation is an integro-differential equation for the interface
coordinate $Y_I(X)$. Boundary conditions need to be imposed at a distance of
order $\xi$ from the boundary. At this scale the shear is negligible so that
the equilibrium order parameter profile is adequate. For unforced
conditions, this leads to $Y_I^\prime=0$, although other situations can be
accommodated.~\cite{forcing} According to Eq.~(\ref{modelB}), the curvature
of the interface satisfies the following scaling relation 
\begin{eqnarray}
R^{-1}(X) = f_1\left(\left(l_0/L\right)^2, X \right) \ .  \label{eq:scalingB}
\end{eqnarray}
Hence, the interface configuration is solely determined by the dimensionless
number $(l_0/L)^2$, and by the equilibrium boundary conditions on the order
parameter at $x=\pm L/2$.

\begin{figure}[p]
\epsfig{figure=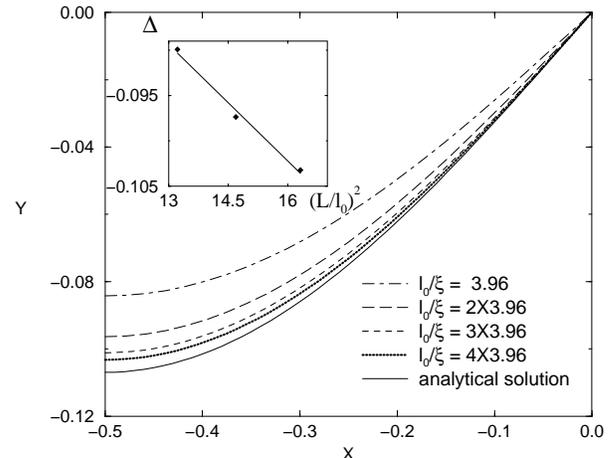,width=3.1in}
\caption{Scaled steady state configuration of a passive interface
for the values of $l_0/\protect\xi$ listed, and $L = 4.04 l_{0}$. 
The inset shows the maximum interface deflection $\Delta$ as a function
of $(L/l_{0})^2$ at $l_{0}/\xi = 4 \times 3.96$.}
\label{fi:noflow}
\end{figure}

Since Eq.~(\ref{modelB}) is expected to be valid only on length scales large
compared to $\xi$, we have numerically solved Eq.~(\ref{Cahn-Hill}) to
verify the above scaling form, as well as to determine the approach to this
scaling form as $l_0/\xi >> 1.$ The details of the numerical algorithm have
been described elsewhere \cite{J&V}. Briefly, the fluid is in a rectangular
domain $-L/2 \le x \le L/2$ and $-3L/2 \le y \le 3L/2$, with boundary
conditions $\hat{{\bf n}} \cdot \nabla \phi = 0$ and $\hat{{\bf n}} \cdot
\nabla (\nabla^{2}\phi )= 0$ at $x = \pm L/2$, and $\phi = \phi_{\pm}$ at $y
= \pm 3L/2.$ (The corresponding equilibrium contact angle is $\pi/2$; the
boundary conditions also ensure no flux of $\phi$ through the lateral
walls.) The interface is initially located at $y=0$, and we integrate 
Eq.~(\ref{Cahn-Hill}) forward in time until a stationary configuration is
reached. The local part of free energy density is the standard double well
potential \cite{J&V}, and a square grid has been used of side no larger than 
$\xi/3$ so that the relative error in the location of the interface is 5\%
or less. Fig.~\ref{fi:noflow} shows the interfacial configuration for 
a range of $l_0/\xi$. The curves tend to superpose 
as $l_0/\xi$ becomes larger for fixed $l_0/L,$
supporting the scaling prediction. 

For the range of parameters considered, the interfacial displacement away
from planarity is small. Therefore we also obtain a perturbative solution of
Eq.~(\ref{modelB}). Direct substitution of the Green's function \cite{green}
into Eq.~(\ref{modelB}) leads to, 
\begin{eqnarray}
Y_{I}(X)=\frac{2}{\pi ^5}\left( \frac{L}{l_{0}}\right) ^{2}
\sum_{n=1}^{\infty }\frac{(-1)^{n-1}}{ (2n-1) ^{5}}\sin \left(
(2n-1)\pi x\right) \ ,  \label{eq:YI}
\end{eqnarray}
at lowest order in $Y_{I}$. Note that the term corresponding to $n=1$
dominates so that the correct dimensionless parameter for the expansion
should be $(L/l_{0})^{2}\pi ^{-5}$. We next define $\delta =-1/2+X$ and find
the curvature of the interface near the boundary,
$R^{-1}(-1/2+\delta )=-R_{w}^{-1}+\frac{1}{4\pi }\left( L/l_{0}\right)
^{2}\times \left[ (3-2\ln \left( \pi/2\right) )\delta ^{2}-2\delta ^{2}\ln
\delta \right] +\dots \ ,$ 
where terms $O((L/l_{0})^{4})$ and $O(\delta ^{4})$ have been neglected. The
constant $R_{w}\approx (0.426/2\pi )(L/l_{0})^{2}$. Note that there is a
weak singularity as $\delta \rightarrow 0$ (a divergent second derivative),
that can be traced back to the factor $(2n-1)^{-5}$ in Eq.(\ref{eq:YI}).
The perturbative solution for $Y_{I}$ is also shown in Fig. 1.
The singularity in the second derivative of the curvature is much too weak 
to be observed directly.

We next consider an active interface and allow for the additional flow
induced by order parameter variations. Boundary conditions for Eq.~(\ref
{Navier-Stokes}) are no-slip at the solid boundaries, ${\bf v}=\pm V_{0}/2 
\hat{{\bf y}}$ at $x=\pm L/2$, and a far field velocity approaching ${\bf 
v_{s}}$. The steady state configuration of $\mu $ is again given by Eq. 
(\ref{inverse-Cahn-Hill}), but ${\bf v}$ needs to be determined
self-consistently. Under the same assumptions used to derive the interface
equation in the case of a passive interface and neglecting the inertial
term, one finds \cite{Bray,OK}, 
\begin{eqnarray}
& &\frac{1}{R(X)}=\left( \frac{l_{0}}{L}\right)^{-2}
\int_{-1/2}^{1/2}dX^{\prime }G({\bf R}_{I},{\bf R}_{I}^{\prime }) \left[
- X^{\prime }+\right.  \nonumber \\
& &\left. \frac{1}{Ca}\frac{dS}{dX^{\prime }}\int_{-1/2}^{1/2}dX^{\prime
\prime }\frac{dS}{dX^{\prime \prime }}n_{i}^{\prime }T_{ij}({\bf R}
_{I}^{\prime }, {\bf R}_{I}^{\prime \prime })n_{j}^{\prime \prime }\frac{1}
{R(X^{\prime \prime })}\right] .  
\label{modelH}
\end{eqnarray}
Repeated indices imply summation, and $T_{ij}$ is the Oseen tensor \cite
{re:pozrikidis92}. A capillary number has been defined as $Ca=\eta
V_{0}/\sigma $, $n_{i}^{\prime }=n_{i}(X^{\prime })$ is the $i$th component
of the unit normal to the interface at $X^{\prime }$, and $dS(X)/dX=\sqrt{
1+(dY_{I}(X)/dX)^{2}}$. This interface equation, in turn, suggests the
following scaling form, 
\begin{eqnarray}
R^{-1}(X)=f_{2}\left( Ca,\left( l_0/L\right) ^{2},X\right) \ .  \label{mdHi}
\end{eqnarray}
In the limit $Ca\rightarrow \infty $, the passive case is recovered 
(Eq.~(\ref{modelB})).

\begin{figure}[p]
\epsfig{figure=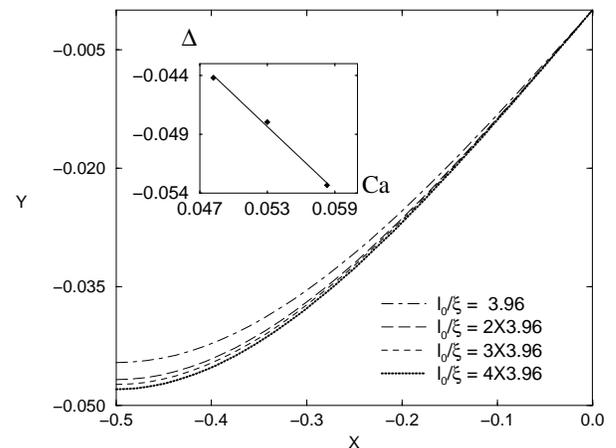,width=3.1in}
\caption{Scaled steady state configuration of an active interface
for the values of $l_{0}/\protect\xi$ listed, $L = 4.04 l_{0}$
with $Ca = 0.053$. The inset reveals the maximum interface deflection
$\Delta$ for different wall velocities (i.e., different $Ca$) 
with fixed $L/\xi$ and  $l_0^2 Ca$.}
\label{fi:flow}
\end{figure}

We have also verified this scaling form by direct numerical solution of
Eqs.~(\ref{Cahn-Hill}) and (\ref{Navier-Stokes}) subject to the boundary
conditions given: no-slip at $x=\pm L/2$, and ${\bf v}={\bf v_{s}}$ at 
$ y=\pm 3L/2$. Figure \ref{fi:flow} shows the results obtained for fixed 
$l_{0}/L$ and $Ca$ and a range of values of $l_{0}/\xi $. 
Again, these results are consistent with the scaling forms derived. Note
that the values of $l_{0}/L$, and $l_{0}/\xi $ are the same in both 
Figs.~\ref{fi:noflow} and \ref{fi:flow}, but the interface deflection in 
Fig.~\ref{fi:flow} is much smaller than that in Fig.~\ref{fi:noflow}. This 
is a consequence of the additional dissipation mechanism in the fluid
and the induced flow. The small values of
$Y_{I}$ in the numerical solution suggest that $Y_{I}$ is proportional
to $V_{0}$, as can be seen from the inset in Fig.~2.

We now compare our results with previous studies of immiscible fluids
based on a macroscopic description~\cite{re:cox86,re:dussan86,re:degennes85}. 
Equation~(\ref{mdHi}) can be written in the following scaling form 
$l_{0}/R_{c}(x/l_{0})=\tilde{f_{2}} (Ca,l_{0}/L,x/l_{0})$. Now choose the
origin such that the walls are located at $x=0$, and $L$. In the limit 
$ x/L\ll 1$ but $x/l_{0}\gg 1$, one expects that the dependence of the
curvature on both $L$ and $l_{0}$ is weak. If the leading behavior is such
that the curvature is independent of both $L$ and $l_{0}$, then the scaling
function reduces to $\tilde{f_{2}}\sim \tilde{f}_{3}(Ca)l_{0}/x$. Using the
fact that $\tilde{f}_{3}(Ca) \simeq A_0 Ca$ \cite{footnote}, this relation
can be integrated to yield $\frac{dy_{I}}{dx}=A_{0} \ Ca\ln \frac{x}{x_{0}}
+B_0$ where $A_{0}$ and $B_0$ are constant, and $x_{0}/l_{0}\gg 1.$ This
\lq\lq outer solution'', valid for small $Ca$ and $x/l_{0}\gg 1$, has the same
functional form as the immiscible case~\cite{re:cox86}. There, the expansion 
parameter in the matched asymptotic expansion
is $\epsilon = s/L$, where $s$ is constant and of molecular dimensions.
In the present case, however, the inner region is determined by $l_{0}$,
which is a function of the boundary velocity. Therefore a similar matched 
asymptotic expansion to derive the macroscopic interfacial profile for a
two phase fluid would lead to a qualitatively different behavior. In
particular, note that as $V_{0}$ becomes small, nonlocal diffusive
coupling between distant regions of the interface becomes important.
We note finally that our conclusions are based on the scaling form
derived from Eq.~(\ref{modelH}), which is valid only for $x \gg \xi$, although
it has also been numerically validated for the range of parameters that
we can explore computationally.

To summarize, we have studied the steady state interfacial configuration in
a partially miscible two-phase fluid under steady shear. Including mutual
diffusion and miscibility naturally introduces a
cutoff, physical in nature, which eliminates stress singularities. An
approximate integro-differential equation for the steady state interface
configuration has been derived, valid at distances larger than the
interfacial thickness $\xi$. For a passive interface,
the configuration satisfies a scaling form that depends only on
the ratio $l_0/L$, where $l_{0} = \sqrt{\xi D/V_{0}}$.
We then argue that the length scale $l_{0} $, which is much larger than
molecular dimensions but not necessarily macroscopic, defines the scale over
which effective interface slip occurs purely due to diffusive 
relaxation of the order parameter. Similar results are obtained for an 
active interface, with the scaling function now depending on both $l_{0}/L$ 
and the capillary number $Ca = \eta V_{0}/\sigma$. 
Numerical solutions of the full dynamical equations, Eqs.~(\ref{Cahn-Hill})-
(\ref{Navier-Stokes}), support both scaling forms.
Our results indicate that the slip mechanism of an interface separating 
coexisting fluids is qualitatively different than in the immiscible
case. Experiments involving very small $Ca$ on coexisting fluids would 
be useful to verify our results.

DJ is grateful for the support of the NSF under DMR9217935. JV is supported
by the U.S. Department of Energy, contract No. DE-FG05-95ER14566, and also
in part by the Supercomputer Computations Research Institute, which is
partially funded by the U.S. Department of Energy, contract No.
DE-FC05-85ER25000.


\begin{references}
\bibitem{forcing}  Introducing wall potentials into the coarse grained free
energy will allow different values of the equilibrium contact angle.
\end{references}
\end{document}